\documentclass[aps,twocolumn]{revtex4}
\usepackage{epsf}
\newcommand{\A}{{\cal{A}}}
\newcommand{\B}{{\cal{B}}}
\newcommand{\N}{{\cal{N}}}
\newcommand{\M}{{\cal{M}}}
\newcommand{\Pe}{{\cal{P}}} 
\newcommand{\R}{{\cal{R}}}
 
\begin{document}
 
\title{Interplay between the single particle and collective \\ 
       features in the boson fermion model}

\author{T.\ Doma\'nski$^{(a,b)}$ and J. Ranninger$^{(a)}$}
\affiliation{
 $^{(a)}$  
           Centre de Recherches sur les Tr\`es Basses Temp\'eratures
           CNRS, 38-042 Grenoble Cedex 9, France \\   
 $^{(b)}$  
           Institute of Physics, M.\ Curie Sk\l odowska University, 
           20-031 Lublin, Poland 
}

\begin{abstract}
We study the interplay between the single particle and fermion pair 
features in the boson fermion model, both above and below the transition 
temperature $T_{c}$, using the flow equation method. Upon lowering
the temperature the single particle fermionic spectral function: 
(a) gradually develops a depletion of the low energy states 
    (pseudogap) for $T^{*}>T>T_{c}$ and a true superconducting 
    gap for $T<T_{c}$, 
(b) exhibits a considerable transfer of spectral weight between 
    the incoherent background and the narrow coherent peak(s) 
    signifying long-lived quasi-particle features.

The cooperon spectral function consists of a  delta function peak, centered 
at the renormalized boson energy $\omega=\tilde{E}_{\bf q}$, and 
a surrounding incoherent background which is spread over a wide 
energy range. When the temperature approaches $T_{c}$ from above this 
peak for ${\bf q}={\bf 0}$ moves to $\omega=0$, so that the static 
pair susceptibility diverges (Thouless criterion for the broken 
symmetry phase transition). Upon  decreasing the temperature 
below $T_{c}$ the cooperon peak becomes the collective (Goldstone) 
mode $E_{\bf q} \propto |{\bf q}|$ in the small momentum region 
and simultaneously  splits off from the incoherent background 
states which are expelled to the high energy sector $|\omega| \geq 
2 \Delta_{sc}(T)$. We discuss the smooth evolution of these features 
upon approaching $T_{c}$ from above and consider its feedback 
on the single particle spectrum where a gradual formation of 
 damped Bogoliubov modes (above $T_{c}$) is observed.
\end{abstract}

  
\maketitle

\section{Introduction}

The boson fermion model (BFM) was initially invented for a
description of the conduction band electrons coupled to the lattice 
vibrations in the region of intermediate electron-phonon coupling 
\cite{Ranninger-85}. The underlying physics emerges from the assumption 
that in the crossover region the electrons exist partly in form of 
bound pairs (hard-core bosons) and partly as quasi free particles 
(fermions). Many-Body correlations of such a two component system are 
induced due to charge exchange processes converting the fermion 
pairs into the hard-core bosons and vice versa. The initially very 
heavy bosons effectively increase their mobility and, at the 
critical temperature $T_{c}$, undergo a Bose Einstein (BE) 
condensation while the fermions are simultaneously driven to 
a superconducting state. 

Upon approaching $T_{c}$ from above there are several precursor 
features of superfluidity/superconductivity showing up in the system. 
In this paper we address these precursor effects by means of a 
renormalization group scheme which is outlined in some detail in 
chapter II. We will show that, {\em in the pseudogap phase ($T<T^{*}$) 
and in superconducting state ($T<T_{c}$), the single- and 
two-particle properties become strongly interdependent}. In 
the pseudogap phase this is seen for instance through a gradual 
destruction of the single particle states (near the Fermi surface)
which is accompanied by a simultaneous emergence of fermion pair 
states. Fermion pairs with total zero momentum show up in the 
normal state only as  damped entities which propagate over a 
finite time and/or spatial scales. However, we find that there 
is a certain critical momentum $q_{crit}(T)$ above which 
fermion pairs become long lived quasi-particles (which are 
separated from the incoherent background as discussed in section 
III). At the phase transition $q_{crit}(T_{c}) \rightarrow 0$ and 
below all the cooperons become good quasiparticles. The appearance 
of fermion pairs above $T_{c}$ affects the single particle 
spectrum, leads  to the formation of the Bogoliubov shadow 
branches as has been discussed by us recently \cite{PRL-03}. 
When  passing through the phase transition at $T_{c}$ we expect 
a smooth evolution from the pseudo- to the fully gaped single 
particle spectrum where the Bogoliubov features (caused by the 
existence of the fermion pairs) are present below as well as 
above $T_{c}$. In the superconducting state we find that 
the interdependence between the single- and two-particle 
properties leads to the characteristic peak-dip-hump structure. 
Similar conclusions have recently been reached independently  
by Pieri et al.\ \cite{Pieri-04} within a different
theoretical approach.

As has been widely emphasized by Uemura \cite{Uemura-03}, there are 
a number of very convincing experimental indications for precursor 
phenomena in the underdoped HTS cuprates. Whether the whole pseudogap 
phase can be exclusively attributed to such precursor effects is still 
under debate. Nevertheless, for temperatures sufficiently close 
to $T_{c}$ (in the underdoped samples) the existence of fermion 
pairs, being correlated on a  small spatial and temporal scale,  
were confirmed by the measurements of the optical conductivity 
in the terahertz regime \cite{Corson-99}. On the other hand
 the static experiments measuring the Nernst coefficient 
 \cite{Ong-00} gave indications for the existence of 
 "moving pairs" with a certain phase slippage above $T_{c}$. 
These facts together with the peak-dip-hump structure found 
by the ARPES \cite{ARPES-exp} acquire  here a natural explanation 
within precursor phenomena which are intrinsic to the BFM or 
similar scenarios, accounting for strong pair fluctuations. 
 
On a more general basis, the BFM is often believed to capture 
essential aspects of the crossover physics between  weakly 
coupled and strongly paired lattice electrons \cite{Leggett-80,
NSR-85}. Various unconventional properties of the superconducting 
state have been investigated within this model by several groups 
\cite{Micnas_etal,Lee_etal,Ranninger_etal,Micnas}. Several 
authors concluded according to phenomenological considerations 
\cite{Enz-96,Geshkenbein-97} that the BFM can serve as an effective 
model for the description of quasi 2-dimensional strongly 
correlated cuprate superconductors \cite{Auerbach-02}.

The crossover issue and the BFM turned out to be of particular 
interest also in  atomic physics \cite{Timmermans,Kokkelmans,
Ohashi}, where a resonant Feshbach scattering is induced between 
the trapped alkali atoms, such as $^{40}$K or $^{6}$Li. By applying 
external magnetic fields the effective interaction between atoms can 
be varied from the weak (BCS) to the strong coupling (BE) limits. 
Under optimal conditions a {\em resonant superconductivity} 
is expected to arise at $T_{c} \sim 0.5 T_{F}$ \cite{Ohashi}, 
which is presently routinely observed in  several research labs 
\cite{Ketterle-03}. This very general scenario of the Feshbach 
resonance can be theoretically expressed via the BFM, as 
was recently shown by one of us \cite{PhysRevA-03}.

It has been frequently stressed in the literature 
\cite{Tchernyshyov-97,Levin,Giovannini-01} that a selfconsistent 
and conserving treatment of single-particle and pair correlations 
has a crucial importance for the description of the HTS cuprates. 
In this paper we study the mutual interdependence between such 
single- and two-particle properties (paying special attention 
to the precursor features) by extending our previous work 
\cite{Domanski-01} based on the {\em flow equations} method 
\cite{Wegner-94}. Our former study  focused on a diagonalization
of the Hamiltonian and the determination of the renormalized fermion 
and boson energies \cite{Domanski-01}. In the present paper we 
derive the Green's functions (dynamic quantities) which determine 
the propagation of single fermions, single bosons and of fermion 
pairs. From these functions we obtain the corresponding excitation 
spectra. The methodological virtue of the flow equations method 
is that, besides treating the single- and two-particle entities 
on equal footing, it distinguishes between the contribution of  
long-lived and damped quasi-particles in the spectrum. The former 
are usually represented by  delta function peaks with a given
spectral weight while the latter are given in form of a broad 
incoherent background. 

For our study we use the following Hamiltonian 
\cite{Ranninger-85}
\begin{eqnarray}
H & = & \sum_{{\bf k},\sigma} \left( \varepsilon_{\bf k} - 
\mu \right) c_{{\bf k}\sigma}^{\dagger} c_{{\bf k}\sigma} 
+ \sum_{\bf q} \left( \Delta_{B} - 2\mu \right) 
b_{\bf q}^{\dagger} b_{\bf q} \nonumber \\ & + & 
\frac{v}{\sqrt{N}} \; \sum_{{\bf k},{\bf q}} \left(  
b_{\bf q}^{\dagger} c_{{\bf q}-{\bf k}\downarrow}
c_{{\bf k}\uparrow} + \mbox{h.c.} \right) \;,
\label{BF}
\end{eqnarray}
where the operators $c_{{\bf k}\sigma}^{\dagger}$ ($c_{{\bf k}\sigma}$) 
refer to creation (annihilation) of fermions with the energy 
$\varepsilon_{\bf k}$ and $b_{\bf q}^{\dagger}$ ($b_{\bf q}$) 
correspondingly to bosons in localized states  $\Delta_{B}$. 
The boson-fermion coupling $v$ will be taken here as isotropic, 
although for the real HTS systems it should be used with a 
$d$ wave prefactor \cite{Micnas,Enz-96,Geshkenbein-97}. For 
simplicity we  neglect here also the hard core property of bosons, 
which is justified as long as the concentration of bosons is small.

\section{The method}

\subsection{Generalities}

We apply a canonical transformation $S(l)$ in order 
to eliminate the interaction between the boson and the 
fermion subsystem. This transformation will be carried 
out in a continuous way ($l$ denotes the continuous
{\em flow parameter}) so, that the transformed
Hamiltonian $H(l)=e^{S(l)}He^{-S(l)}$ reduces 
to a manageable form for further analysis. The more
generally known classical single step transformations 
projects out the terms which are linear with respect 
to a given perturbation. Here we demand much more 
stringent constraints on a transformed Hamiltonian 
going beyond such a standard perturbative scheme.
  
The evolution of the Hamiltonian $H(l)$ with 
respect to the varying flow parameter $l$ 
is determined through the  differential equation
\begin{eqnarray}
dH(l)/dl = [ \eta(l),H(l) ] 
\label{Hflow}
\end{eqnarray}
subject to the initial condition $H(0)=H$. A generating
operator is defined by $\eta(l) \equiv \left( de^{S(l)}/dl 
\right) e^{-S(l)}$.

In principle, one can transform the Hamiltonian in many 
different ways by choosing various operators $\eta(l)$ 
(or $S(l)$). Some particularly efficient schemes have been 
proposed by Wegner  \cite{Wegner-94} and independently 
by Wilson and G\l azek \cite{Wilson-94} going back to 
the RG approach ideas \cite{RG-idea}. Through a continuous 
transformation of the Hamiltonian one effectively renormalizes 
its {\em coupling constants} while keeping a given constrained 
structure. In other words, the parameters of the Hamiltonian 
such as the energies, the two-body potentials and so on are 
assumed to be $l$-dependent.

In some distinction from the RG approach one does not
integrate out the high energy states, but instead of it 
they are renormalized in the initial part of transformation 
until $l \sim (\Delta \varepsilon)^{-2}$ \cite{Wegner-01}. 
Subsequently, states with small energy differences start 
to be renormalized and finally, for $l \rightarrow \infty$, 
the transformed Hamiltonian $H(\infty)$ eventually 
reduces to a (block) diagonal structure \cite{Mielke-98}.

\subsection{Implementation to BF model}

In our previous work \cite{Domanski-01} we have derived the 
continuous canonical transformation for block diagonalization
of the BFM. The transformed Hamiltonian was con\-strained 
to the following structure $H(l)=H_{0}(l)+H_{int}(l)$, 
where
\begin{eqnarray}
H_{0}(l) & = & \sum_{{\bf k},\sigma} (\varepsilon_{\bf k}(l)
- \mu) c_{{\bf k}\sigma}^{\dagger}c_{{\bf k}\sigma} +
\sum_{\bf q} (E_{\bf q}(l)-2\mu) b_{\bf q}^{\dagger}b_{\bf q} 
\nonumber \\ & + & \frac{1}{N}
\sum_{{\bf k},{\bf p},{\bf q}} U_{{\bf k},{\bf p},{\bf q}}(l)
c_{{\bf k}\uparrow}^{\dagger} c_{{\bf p}\downarrow}^{\dagger}
 c_{{\bf q}\downarrow} c_{{\bf k}+{\bf p}-{\bf q}\uparrow} \;,
\label{H_0} \\   
H_{int}(l) & = & \frac{1}{\sqrt{N}}\sum_{{\bf k},{\bf p}} 
v_{{\bf k},{\bf p}}(l) \left(  b_{{\bf p}+{\bf k}}^{\dagger}
c_{{\bf k}\downarrow} c_{{\bf p}\uparrow} +
\mbox{h.c.} \right) \;.
\label{H_pert}
\end{eqnarray}
For eliminating $H_{int}(l)$ we follow
the idea proposed by Wegner \cite{Wegner-94} who  
showed that, upon using $\eta(l) = [H_{0}(l),H_{int}(l)]$, 
one obtains $\lim_{l\rightarrow\infty}H_{int}(l)=0$.
In this case
\begin{eqnarray}
\eta(l)= - \frac{1}{\sqrt{N}} \sum_{{\bf k},{\bf p}}
\alpha_{{\bf k},{\bf p}}(l) \left(  b_{{\bf p}+{\bf k}}^{\dagger}
c_{{\bf k}\downarrow} c_{{\bf p}\uparrow} - 
\mbox{h.c.} \right) ,
\label{eta}
\end{eqnarray}
where $\alpha_{{\bf k},{\bf p}}(l)=\left( \varepsilon_{\bf k}(l)
+\varepsilon_{\bf p}(l)-E_{{\bf k}+{\bf p}}(l)\right) v_{{\bf k},
{\bf p}}(l)$. All $l$-dependent parameters of the Hamiltonian 
(\ref{H_0},\ref{H_pert}) are determined via a set of the flow 
equations (16-21) given in Ref.\ \cite{Domanski-01}. They are 
obtained from the operator equation (\ref{Hflow}) by reducing 
the higher order interactions through normal ordering 
(linearization).

Since at $l=\infty$ the bosons are no longer hybridized with
fermions, we essentially obtain the (semi) free subsystems 
with renormalized effective spectra. Only the fermion part 
contains the long range Coulomb interaction $U _{{\bf k},
{\bf p},{\bf q}}$ which in some cases can play an  important role.
For instance, in the ${\bf q}={\bf p}$ channel one obtains 
\cite{PhysRevA-03} a resonant type amplitude of the potential 
$U _{{\bf k},{\bf p},{\bf p}}(\infty)$ for $\varepsilon_{\bf k}
+\varepsilon_{\bf p} = E_{{\bf p}+{\bf k}}$. This corresponds 
to the resonant (Feshbach) scattering between electrons when 
their total energy is equal to energy of the bound pair 
(hard-core boson) \cite{Timmermans}. In the context of HTS 
such unusual scattering is also important leading to the 
particle-hole asymmetry of the low energy spectrum both, 
in the pseudogap and superconducting phases \cite{PhysicaC-03}.

One should  remember that the amplitude of the induced Coulomb 
potential $U_{{\bf k},{\bf p},{\bf q}}(\infty)$ is finite for 
any channel. It was estimated to be residual, of the order 
$v^{2}$ \cite{Domanski-01}. In the following we will thus 
treat the transformed Hamiltonian
\begin{eqnarray}
H(\infty)=H^{F}(\infty)+H^{B}(\infty) 
\end{eqnarray}
as composed of two contributions from bosons  
$H^{B}(\infty) = \sum_{\bf q} \tilde{E}_{\bf q}
b_{\bf q}^{\dagger}b_{\bf q}$ with their effective
energy $\tilde{E}_{\bf q} \equiv E_{\bf q}(\infty)-2\mu$,
and fermions approximated by
$H^{F}(\infty) \simeq \sum_{{\bf k}\sigma}
\tilde{\varepsilon}_{\bf k} 
c_{{\bf k}\sigma}^{\dagger}c_{{\bf k}\sigma}$ 
with the effective energy
\begin{eqnarray}
\tilde{\varepsilon}_{\bf k} \equiv
\varepsilon_{\bf k}(\infty) - \mu + \frac{1}{N} 
\sum_{\bf p} U _{{\bf k},{\bf p},{\bf p}}(\infty) 
n_{{\bf p},-\sigma}^{F} .
\end{eqnarray}
Here $n_{{\bf p},\sigma}^{F}=\langle c_{{\bf p},\sigma}
^{\dagger}c_{{\bf p},\sigma}\rangle$, where the spin is 
a dummy index which will be kept throughout the remainder 
of this paper in order to indicate the origin of such terms.

\subsection{Dynamical quantities}

In this work we focus on determining  thermal 
equilibrium Green's functions of the form
\begin{eqnarray}
\langle \langle O_{1} ( \tau ) ; O_{2} \rangle \rangle = 
- \; \langle {\cal{T}}_{\tau} O_{1}(\tau) O_{2} \rangle
\label{GF}
\end{eqnarray}
where the time evolution of the operators is given by 
$O(\tau)=e^{\tau H}Oe^{-\tau H}$, with $\tau \in
\left< 0, \beta \right>$ and $\beta=1/k_{B}T$.
As usual, ${\cal{T}}$ denotes ordering with 
respect to the imaginary time $\tau = it$.

The computation of the thermal averages $\langle ... \rangle 
= \mbox{Tr} \left\{ e^{-\beta H} ... \right\} / \mbox{Tr}
\left\{ e^{-\beta H} \right\}$ is easiest to carry out 
using the transformed Hamiltonian $H(\infty)$ because 
of its (block-) diagonal structure.  Due to the invariance 
of the trace under the unitary transformation we can write
\begin{eqnarray}
\mbox{Tr} \left\{  e^{-\beta H} O \right\} & = &
\mbox{Tr} \left\{ e^{S(l)}  e^{-\beta H} O e^{-S(l)} 
\right\} \nonumber \\ & = & 
\mbox{Tr} \left\{  e^{-\beta H(l)} O(l) \right\}  \;, 
\label{trace}
\end{eqnarray}
where $O(l)=e^{S(l)}Oe^{-S(l)}$. Hence, if we want 
to use the transformed Hamiltonian $H(l=\infty)$ in 
the Boltzmann factor $e^{-\beta H(l)}$ we ought to 
transform the observable $O$ too. For 
the continuous transformation this is however a nontrivial 
problem because, in order to get $O(\infty)$, one must 
analyze the whole transformation process. The evolution 
of the arbitrary observable $O(l)$ with respect to $l$ 
must be deduced on a basis of the differential equation
\begin{eqnarray}
dO(l)/dl = [ \eta(l),O(l) ] .
\label{flow}
\end{eqnarray}
For the Hamiltonian $O=H$ it thus is given by (\ref{Hflow}) 
which was already discussed previously by us \cite{Domanski-01}
for this model.

In the next sections we study the $l$-dependence of the 
individual boson and fermion operators and of fermion 
pair operators. By looking at the limit $l \rightarrow 
\infty$, we shall derive effective spectral functions
\begin{eqnarray}
A^{F,B,pair}({\bf k},\omega)  = - \frac{1}{\pi} \; 
\mbox{Im}  \; G^{F,B,pair}({\bf k},\omega)  ,
\label{spectral}
\end{eqnarray}
where $G({\bf k},\omega) = \int_{0}^{\beta} d\tau \; 
e^{\tau \omega} G({\bf k},\tau)$ with the following 
single particle Green's functions
\begin{eqnarray}
G^{B}({\bf q},\tau) & = & \langle \langle b_{\bf q} (\tau) ; 
b_{\bf q}^{\dagger}  \rangle \rangle, \label{G_B_def} \\
G_d^{F}({\bf k},\tau) & = &  \langle \langle 
c_{\bf k \uparrow} (\tau) ; c_{\bf k \uparrow}^{\dagger} 
\rangle \rangle, \\
\label{G_F_def}
G_{od}^{F}({\bf k},\tau) & = &  \langle \langle 
c_{\bf k \uparrow} (\tau) ; c_{-{\bf k} \downarrow} 
\rangle \rangle,
\end{eqnarray}
and the two-particle pair propagator given by
\begin{eqnarray}
G^{pair}({\bf q},\tau) = \frac{1}{N^{2}} \sum_{{\bf k},{\bf p}}  
\langle \langle c_{{\bf k} \downarrow} (\tau)
c_{{\bf q}-{\bf k}\uparrow}(\tau); 
c_{{\bf q}-{\bf p}\uparrow}^{\dagger}
c_{{\bf p} \downarrow}^{\dagger} \rangle \rangle .
\label{G_pair}
\end{eqnarray}
We will next investigate the structure of these spectral 
functions and discuss their related physical properties.

\section{Bosons and cooperons}

\subsection{Flow of the boson operators}

In the course of such a continuous transformation, 
the initial boson operator $b_{\bf q}$ becomes convoluted 
for $l>0$ with the fermion pair (cooperon) operator. This 
can be seen from the $l=0$ derivative
\begin{eqnarray}
\frac{db_{\bf q}}{dl} = [\eta(0),b_{\bf q}] = \frac{1}
{\sqrt{N}} \sum_{\bf k} \alpha_{{\bf k},{\bf q}-{\bf k}}(0)
c_{{\bf k}\downarrow} c_{{\bf q}-{\bf k}\uparrow}. 
\label{b_indication}
\end{eqnarray}
Physically this means, that while disentangling the boson 
from fermion subsystem we obtain some new quasiparticles 
made out of the initial bosons and cooperons (like in the BCS 
theory where the quasi-particles are composed of electrons and 
holes).

Guided by the structure of  to (\ref{b_indication}) it is judicious  
to choose the following superposition for the $l$-dependent boson 
operator	
\begin{eqnarray}
b_{\bf q}(l)= \A_{\bf q}(l)  b_{\bf q} + 
\frac{1}{\sqrt{N}} \sum_{\bf k} \B_{{\bf q},{\bf k}}(l)
c_{{\bf k}\downarrow} c_{{\bf q}-{\bf k}\uparrow},
\label{b_Ansatz}
\end{eqnarray}
where $\A_{\bf q}(l)$, $\B_{{\bf q},{\bf k}}(l)$ are 
some complex functions with the initial condition 
$\A_{\bf q}(0)=1$ and $\B_{{\bf q},{\bf k}}(0)=0$. 
Substituting (\ref{b_Ansatz}) into the flow equation 
(\ref{flow}) we obtain 
\begin{eqnarray}
\frac{d\A_{\bf q}(l)}{dl} & = &  - \frac{1}{N} 
\sum_{\bf k} \alpha_{{\bf k},{\bf q}-{\bf k}}(l)
\; f_{{\bf k},{\bf q}-{\bf k}} \;
\B_{{\bf q},{\bf k}}(l) ,
\label{A_flow}
\\
\frac{d\B_{{\bf q},{\bf k}}(l)}{dl} & = & 
\alpha_{{\bf k},{\bf q}-{\bf k}}(l)  \A_{\bf q}(l) ,
\label{B_flow}
\end{eqnarray}
where we introduced the shorthand notation
\begin{eqnarray} 
f_{{\bf k},{\bf p}} = 1 - n_{{\bf k}\downarrow}^{F}   
- n_{{\bf p}\uparrow}^{F} .
\end{eqnarray}
From (\ref{A_flow},\ref{B_flow}) we  notice 
the following invariance 
\begin{eqnarray}
| \A_{\bf q}(l)|^{2} + \frac{1}{N} 
\sum_{\bf k} | \B_{{\bf q},{\bf k}}(l) |^{2}
f_{{\bf k},{\bf q}-{\bf k}} = 1
\end{eqnarray}
which guarantees that the commutation relations 
between the $l$-dependent boson operators
$\left[ b_{\bf q}(l), b_{\bf p}^{\dagger}(l)
\right]=\delta_{{\bf q},{\bf p}}$ are correctly 
preserved.

The parameterization (\ref{b_Ansatz}) which follows from 
the flow equation (\ref{flow}) for the operator $b_{\bf q}(l)$ 
yields the following boson spectral function (\ref{spectral})
\begin{eqnarray}
A^{B}({\bf q},\omega) & = & | \A_{\bf q}(\infty)|^{2} 
\delta \left( \omega - \tilde{E}_{\bf q} \right)
\label{b_spectral} \\
& + & \frac{1}{N} \sum_{\bf k} f_{{\bf k},{\bf q}-{\bf k}}
| \B_{{\bf q},{\bf k}} (\infty) |^{2} 
\delta \left( \omega - \tilde{\varepsilon}_{\bf k}
- \tilde{\varepsilon}_{{\bf q}-{\bf k}} \right). 
\nonumber
\end{eqnarray}
The first term in equation (\ref{b_spectral}) describes 
the  coherent part of the boson spectral function corresponding 
to the long-lived quasi-particles with the renormalized energy 
$\tilde{E}_{\bf q}$. The second contribution describes some 
incoherent background of the boson spectral function which 
represents the states of a relatively short life-time.

We solve  the flow equations (\ref{A_flow},\ref{B_flow}) fully 
selfconsistently applying the numerical procedure based on the 
Runge-Kutta algorithm. For any value of $l$ we discretized 
the coefficients $\A_{\bf q}(l)$, $\B_{{\bf q},{\bf k}}(l)$
using a mesh of 4000 equidistant points for representing the 
vectors ${\bf k}$ and ${\bf q}$ in the Brillouin zone. Due to 
computational limitations we restrict ourselves to a bare
one-dimensional tight binding dispersion $\varepsilon_{\bf k}
(l=0)=-2t \mbox{cos}(k_{x}a)$  and throughout this paper 
 use the bandwidth $D=4t$ as a unit for energies and 
for the  temperature. Starting from the initial value 
$\A_{\bf q}(0)=1$ the $l$-dependent coefficients are calculated 
via the following scheme $\A_{\bf q}(l+\delta l) = \A_{\bf q}(l)
+\delta l \frac{d\A_{\bf q}(l)}{dl}$, where the derivative
is given in equation (\ref{A_flow}). The coefficients
$\B_{{\bf q},{\bf k}}(l)$  are determined in the same way. 
Since the renormalizations of both these coefficients (as well 
as other quantities such as energies $\varepsilon_{\bf k}(l)$, 
$E_{\bf q}(l)$ and boson-fermion coupling $v_{{\bf k},{\bf p}}(l)$) 
occur at the initial steps  of the transformation procedure, we adjust 
the increment $\delta l$ in the following way: $\delta l=0.01$ 
(for $l \leq 5$), $\delta l= 0.1$ (for $5 < l \leq 10^{2}$), 
$\delta l = 1.0$ (for $ 10^{2} < l \leq 10^{3}$), and $\delta l=10$ 
(for $10^{3} < l \leq 10^{4}$), where both $l$ and $\delta l$ 
are expressed in units $D^{-2}$. The asymptotic (fixed points)
are obtained  already around $l \simeq 500$ but the transformation 
procedure is continued up to a good convergence i.e., $l=10^{4}$.

\begin{figure}
\epsfxsize=7.5cm
\centerline{\epsffile{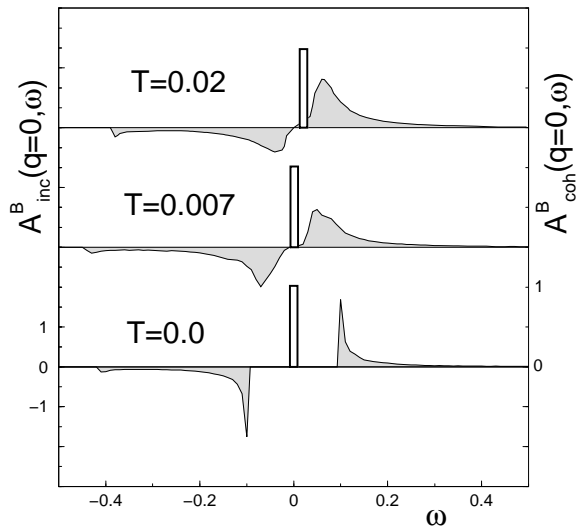}}
\caption{The boson spectral function
$A^{B}({\bf q}={\bf 0},\omega)$ for three  
representative temperatures $T=0.02 > T^{*}$
(top), $T=0.007 < T^{*}$ (middle) and 
for the ground state superconducting phase
$T=0$ (bottom). The BFM parameters are
$n_{tot}=1$, $v=0.1$, $\Delta_{B}=-0.6$
(energies are expressed in units of the initial
fermion bandwidth $D \equiv 1$).
}
\label{Fig1}
\end{figure}

Figure \ref{Fig1} shows the results obtained numerically for 
the single particle boson spectral function $A^{B}({\bf q},
\omega)$ in the long wavelength limit ${\bf q}={\bf 0}$. 
We illustrate three distinct situations corresponding to: 
the normal phase above $T^{*}$ (top panel), the normal phase 
with the pseudogap structure present in the single particle 
fermion spectrum (middle panel) and the superconducting state 
at $T=0$ (bottom panel). In this paper we have chosen the 
same set of parameters as previously \cite{Domanski-01}, 
i.e.\ $\Delta_{B}=0.4$, $n_{tot}=1$, $v=0.1$ such, that the 
temperature at which the pseudogap begins to open up is 
roughly $T^{*} \sim 0.1$.

Our study of the superconducting phase (following the previous 
work \cite{Domanski-01}) is based on a three-dimensional system 
with a BCS type of approach, as far as the fermionic subsystems 
is concerned, and a BE condensation approach for free bosons, as 
far as the bosonic subsystem is concerned. We notice, that in 
this phase there is a perfect separation of the coherent part 
(describing the long lived quasiparticles) from the incoherent 
part of the spectrum. Moreover:
(a) the coherent peak is pinned at $\omega=0$ allowing 
    for a macroscopic occupancy of the zero momentum 
    state by a certain fraction $n_{cond}^{B}$ of 
    the BE condensed bosons,
(b) the incoherent part $A_{inc}^{B}({\bf q},\omega)$ 
    exists only outside a energy window (equal $ 2 v \;
    \sqrt{n_{cond}^{B}}$ as will be explained in section IV.D). 
Owing to such a behavior, the condensed bosons are not damped 
and they are able to establish a long range order parameter 
in the boson subsystem. On the other hand in the normal phase 
above $T^{*}$ the coherent and incoherent parts overlap with
each other and consequently the boson quasiparticles are damped. 
This damping is caused by some very reduced remanent inter-boson 
interaction of the order of  $v^4$, which arises in this 
renormalization procedure \cite{Domanski-01}. 

The pseudogap phase (middle panel of Fig.\ \ref{Fig1}) represents 
some intermediate situation, where we notice that the incoherent 
background is partly pushed away from the coherent peak. Thus 
the zero momentum bosons start to emerge as better and better 
quasiparticles upon approaching $T_{c}$ from above. Yet, the 
zero momentum boson state is macroscopically occupied only 
below $T_{c}$. 

In the BFM there is a strict relation between the single
particle boson and fermion pair excitation spectra (see
equations (\ref{identity_1},\ref{identity_2}) in the next 
section). By inspecting Fig.\ 1 (and figure 3 presented 
below) we conclude that  zero momentum fermion pairs 
gradually emerge in the pseudogap phase ($T^{*}>T$). Upon 
lowering the temperature, the surrounding incoherent 
background fades away and thus effectively leads to increase 
of the life-time of the zero momentum bosons and fermion 
pairs. For $T<T_{c}$, these entities acquire an infinite 
life-time. Experiments, sensitive to the short lived Cooper 
pairs, should be able to detect their presence above $T_{c}$. 
This type of a precursor phenomenon was indeed observed for 
the HTS cuprates using the alternating magnetic fields in 
the terahertz frequencies regime \cite{Corson-99}. A  residual 
Meissner effect was seen there up to nearly 25 K above the 
transition temperature $T_{c}$ and which is an indication 
that propagating fermion pairs exist there on a corresponding 
short time scale.

\begin{figure}
\epsfxsize=8.5cm\centerline{\epsffile{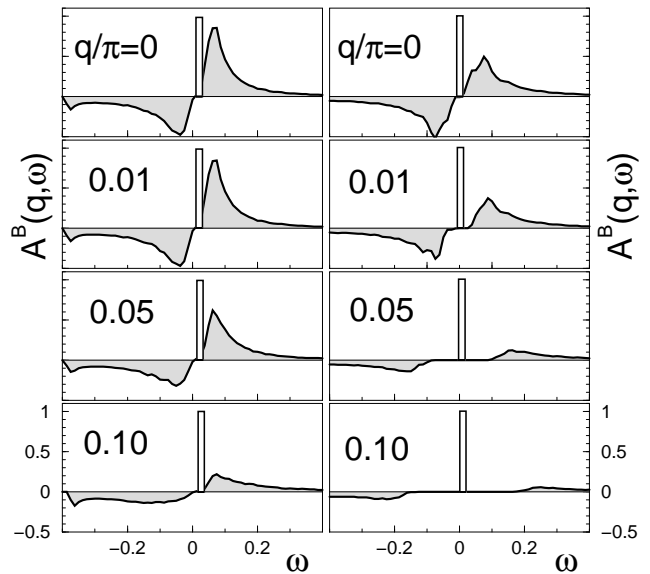}}
\caption{Variation of the boson spectral function 
$A^{B}({\bf q},\omega)$ with respect to changing 
momentum $q$ at high temperature $T=0.02>T^{*}$ 
(panels on the l.h.s.) and low temperature $T=0.004$
(panels on the r.h.s.) corresponding to the pseudogap 
region $T^{*}> T > T_{c}$. The lattice constant is taken 
as a unit $a \equiv 1$.}
\label{Fig2}
\end{figure}

In figure \ref{Fig2} we compare the boson spectral function 
$A^{B}({\bf q},\omega)$ for several momenta $q=|{\bf q}|$ 
at two different temperatures: above (column on the left) 
and below $T^{*}$ (column on the right).  We notice that 
at finite momenta there is is a qualitative difference 
between these spectra. At temperatures $T>T^{*}$ the coherent 
peak is always covered by the incoherent background signaling 
that these components are convoluted with each other. On 
the contrary, at temperatures below $T^{*}$, the coherent 
boson peak separates from the incoherent states. Such a 
splitting-off is very sensitive to moderate temperature changes 
and, for example at $T=0.004$, occurs above a  critical momentum 
$q_{crit} \simeq 0.01 \times \pi/a$. This critical value decreases 
with decreasing temperature and finally $q_{crit}(T_{c}) 
\rightarrow 0$. We thus see that for momenta $q>q_{crit}
(T)$ the coherent boson states, representing the long 
lived propagating modes are not scattered by the 
incoherent background.

Our finding, that under certain conditions the boson peak 
becomes well separated from the background of incoherent states 
$A^{B}_{inc}({\bf q},\omega)$, means that in the pseudogap 
regime ($T^{*} > T > T_{c}$) the finite momentum bosons 
are well defined quasi-particles with infinite life-time.
Since bosons are closely related to the fermion Cooper 
pairs, we further conclude that for $q>q_{crit}(T)$ there
exist infinite life-time cooperons above the transition 
temperature $T_{c}$. According to the Heisenberg principle,  
experiments measuring the pair-pair correlations restricted 
to spatial distance smaller than $\Delta x \sim 1 / 
q_{crit}(T)$ (of the order of several lattice constants) 
should be able to detect such fermion pairs.  And indeed, 
such long-lived "moving" fermion pairs were observed up 
to very high temperatures above $T_{c}$ by applying 
a thermal gradient and measuring the Nernst coefficient 
in underdoped HTS materials \cite{Ong-00}.

\subsection{Flow of the cooperon operators}

We have seen in the previous section that the boson operators 
get mixed for $l>0$ with the fermion pair operators.
Now we expect that, in turn, also the latter become 
convoluted  with $b_{\bf q}$ during such a  transformation. 
By calculating the initial ($l$$=$$0$) derivative 
of the cooperon operators $S_{\bf q}^{-} \equiv 
N^{-1} \sum_{\bf k} c_{{\bf k} \downarrow} 
c_{{\bf q}-{\bf k}\uparrow}$ we obtain
\begin{eqnarray}
\frac{d S_{\bf q}^{-}}{dl} = -\frac{1}{\sqrt{N}} 
\sum_{\bf k} \alpha_{{\bf k},{\bf q}-{\bf k}}(0)
 f_{{\bf k},{\bf q}-{\bf k}} \; b_{\bf q} .
\end{eqnarray}
In accordance with the previous substitution 
(\ref{b_Ansatz}) we propose the following 
Ansatz
\begin{eqnarray}
S_{\bf q}^{-}(l) = \frac{1}{\sqrt{N}} \sum_{\bf k} 
\M_{{\bf k},{\bf q}}(l) c_{{\bf k}\downarrow} 
c_{{\bf q}-{\bf k}\uparrow} + \N_{\bf q}(l)
b_{\bf q} 
\label{S_Ansatz}
\end{eqnarray}
with the initial conditions $\M_{{\bf q},
{\bf k}}(0)=1$ and $\N_{\bf q}(0)=0$.
After substituting the expression (\ref{S_Ansatz}) 
into (\ref{flow}) we get  
\begin{eqnarray}
\frac{d\M_{{\bf q},{\bf k}}(l)}{dl} & = & 
\alpha_{{\bf k},{\bf q}-{\bf k}}(l)  \N_{\bf q}(l)
\label{M_flow}
\\
\frac{d\N_{\bf q}(l)}{dl} & = &  - \frac{1}{N} 
\sum_{\bf k} \alpha_{{\bf k},{\bf q}-{\bf k}}(l)
f_{{\bf k},{\bf q}-{\bf k}} \;
\M_{{\bf q},{\bf k}}(l) 
\label{N_flow}
\end{eqnarray}

We recognize that the flow equations (\ref{M_flow},
\ref{N_flow}) for the unknown coefficients $M_{{\bf q},
{\bf k}}(l)$ and $N_{\bf q}(l)$ have a 
structure identical to that given by  (\ref{A_flow},\ref{B_flow}).
There is only a difference in the initial conditions 
which in this case lead to the following invariance 
\begin{eqnarray}
| \N_{\bf q}(l)|^{2} + \frac{1}{N} 
\sum_{\bf k} f_{{\bf k},{\bf q}-{\bf k}}
| \M_{{\bf q},{\bf k}}(l) |^{2}
= 1 - n_{F} .
\label{MN_inv}
\end{eqnarray}
In the r.h.s.\ of equation (\ref{MN_inv}) we made use 
of the property that $N^{-1} \sum_{\bf k} f_{{\bf k},
{\bf q}-{\bf k}} =1 - n_{F}$, where $n_{F}= N^{-1}  
\sum_{{\bf k},\sigma} n_{{\bf k} \sigma}^{F}$ denotes 
the total concentration of fermions. Equation (\ref{MN_inv}) 
assures the proper statistical relation between the cooperon 
operators $\left[ S_{\bf q}^{-}(l),S_{\bf q}^{+}(l) \right]
= N^{-1} \sum_{\bf k} (1 - c_{{\bf q}-{\bf k} 
\uparrow}^{\dagger} c_{{\bf q}-{\bf k}\uparrow} - 
c_{{\bf k}\downarrow}^{\dagger} c_{{\bf k}\downarrow})$,
which can be approximated by the c-number $\simeq 1-n_{F}$.  

With the Ansatz (\ref{S_Ansatz}) and using its hermitian 
conjugate $S_{\bf q}^{+}(l)$ we can now determine the two 
particle fermion Green's function $G^{pair}({\bf k},\tau)$ 
defined in (\ref{G_pair}). The corresponding spectral 
function (\ref{spectral}) becomes
\begin{eqnarray}
& & A^{pair}({\bf q},\omega)  =  | \N_{\bf q}(\infty)|^{2}
\delta \left( \omega - \tilde{E}_{\bf q} \right)
\label{S_spectral} \\
& & + \frac{1}{N} \sum_{\bf k} f_{{\bf k},{\bf q}-{\bf k}}
| \M_{{\bf q},{\bf k}} (\infty) |^{2}
\delta \left( \omega - \tilde{\varepsilon}_{\bf k}
- \tilde{\varepsilon}_{{\bf q}-{\bf k}} \right) .
\nonumber
\end{eqnarray}
The cooperon spectral function turns out to have a structure 
similar to $A^{B}({\bf q},\omega)$, expressed in (\ref{b_spectral}).
This is a general feature of the BFM which, in particular, 
implies that bosons BE condense simultaneously with 
the fermion pairs driven to superconductivity \cite{Kostyrko-96}. 

\begin{figure}
\epsfxsize=8.5cm
\centerline{\epsffile{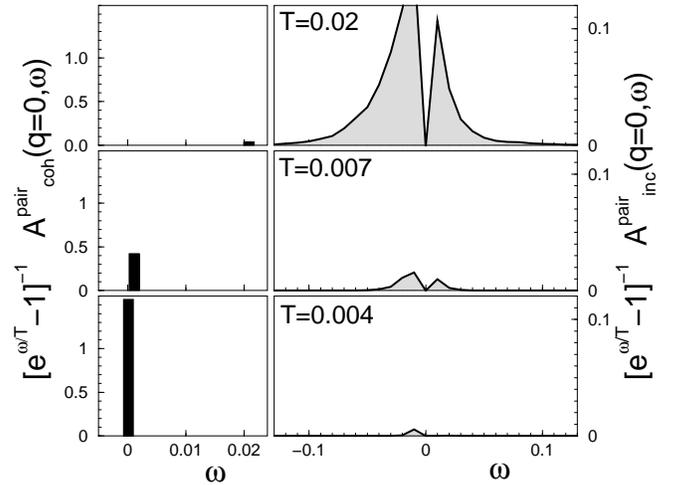}}
\caption{The cooperons' occupancy $\left[ \mbox{exp}(\omega/k_{B}T)
-1\right]^{-1}\times A^{pair}({\bf q},\omega)$ of the zero momentum
coherent (left h.s.\ panel) and incoherent (right h.s.\ panel) states 
for three representative temperatures: $T=0.02$ (above $T^{*}$) 
and $T=0.007$, $0.004$ (below $T^{*}$). With lowering temperature
the incoherent fermion pairs are converted into the coherent
quasi-particles.}
\label{Fig3}
\end{figure}

We limit our quantitative discussion of the cooperon spectral 
function (\ref{S_spectral}) by only presenting in Fig.\ \ref{Fig3}
a distribution of the occupied coherent and incoherent states 
in the long wavelength limit ${\bf q}={\bf 0}$. Above $T^{*}$ 
almost all the cooperons exist as  incoherent objects. Upon 
decreasing the temperature more and more incoherent pairs are 
converted into  coherent ones. Yet, because of their overlap
with the incoherent background the Cooper pairs can propagate 
only on a short time scale.

The fact that the overall structure of the cooperon spectral function 
$A^{pair}({\bf q},\omega)$ is to great extent similar to the single 
particle boson function $A^{B}({\bf q},\omega)$ is not surprising. 
Using for instance the equation of motion technique for the Green's 
functions one can prove the following important identity 
\cite{Kostyrko-96}
\begin{eqnarray}
G^{B}({\bf q},\omega) & = &  G^{B}_{0}({\bf q},\omega)  
\label{B-pair} + \label{identity_1} \\ & + & v^{2} \; 
G^{B}_{0}({\bf q},\omega)  G^{pair}({\bf q},\omega)  
G^{B}_{0}({\bf q},\omega) ,
\nonumber
\end{eqnarray}
where $G^{B}_{0}({\bf q},\omega)=\left[\omega - \Delta_{B} 
+ 2\mu \right]^{-1}$ denotes the single particle boson Green's 
function of the non-interacting system. Both Green's functions 
$G^{pair}({\bf q},\omega)$ and $G^{B}({\bf q},\omega)$ have 
thus common poles and their spectral functions are correspondingly 
related via
\begin{eqnarray}
A^{pair}({\bf q},\omega)  =  \frac{ \left( \omega - \Delta_{B}
+ 2 \mu \right)^{2} }{v^{2}}  \; A^{B}({\bf q},\omega) .
\label{identity_2}
\end{eqnarray}
For ${\bf q}={\bf 0}$, when the single particle boson Green's 
function develops a pole at $\omega=0$, the two particle 
cooperon propagator is characterized by the same pole -  
although with a different weight. This automatically causes 
a divergence of the static pair susceptibility and via the  
Thouless criterion leads to the phase transition into 
a superconducting state.

We want to  stress that all our conclusions concerning the evolution 
of the boson spectral function $A^{B}({\bf q},\omega)$ are also valid 
for $A^{pair}({\bf q},\omega)$. In particular, we emphasize, that in the 
pseudogap region $T^{*} > T > T_{c}$ there exist 
fermion Cooper pairs. Fermion pairs with total zero momentum are 
convoluted with incoherent background states and therefore their
life time is finite. They can be detected experimentally only via 
some short time impulses. On the other hand, long-lived 
fermion pairs can safely exist if their total momentum exceeds 
the critical value $q>q_{crit}(T)$. In this case correlations 
between the fermion pairs can possibly extend over the spatial 
distance up to $\Delta x \sim 1/q_{crit}$. A possibility to 
observe correlations between the fermion pairs above $T_{c}$ 
over a short spatial and temporal scale had been previously 
suggested by Tchernyshyov \cite{Tchernyshyov-97}.

\section{Fermions}

\subsection{Flow of the single particle fermion operators}

Some aspects of the effective single particle fermion 
excitations were already discussed in our recent Letter 
\cite{PRL-03}. We focused our discussion there on the single 
particle fermion spectrum for a narrow region around 
the Fermi surface ${\bf k} \sim {\bf k}_{F}$. We have 
shown that the pseudogap is accompanied by an appearance 
of the Bogoliubov-like branches. The shadow part of these 
Bogoliubov modes is visible above $T_{c}$ as a broad 
structure which narrows upon approaching $T_{c}$. Below 
$T_{c}$ both branches become infinitely narrow, signaling
that bogoliubons become then the true, long-lived quasi-particles. 
This result is herewith confirmed from our present direct study 
of the boson and fermion pair spectra, illustrated in figures 1-3. 
 
In this section we would like to present a detailed derivation 
of the diagonal and off-diagonal parts (in a Nambu spinor 
representation) of the single particle fermion Green's function. 
We will discuss the diagonal and off-diagonal parts studying 
their structure over the whole Brillouin zone above and 
below $T_{c}$. 
 
Let us briefly recapitulate the main properties of the 
fermion operators $c_{{\bf k}\sigma}^{\dagger}$ and 
$c_{{\bf k}\sigma}$ resulting from the continuous canonical
transformation. At the initial step (i.e.\ for $l=0$) their 
derivatives read
\begin{eqnarray}
\mp \frac{d c_{{\bf k}\sigma}}{dl} & = & 
  \alpha_{-{\bf k},{\bf k}}(0) 
  \frac{b_{\bf 0}}{\sqrt{N}} \; c_{-{\bf k},-\sigma}^{\dagger}
\nonumber \\ & + &
  \frac{1}{\sqrt{N}} \sum_{{\bf q} \neq {\bf 0}}
  \alpha_{{\bf q}-{\bf k},{\bf k}}(0)
b_{\bf q} c_{{\bf q}-{\bf k},-\sigma}^{\dagger}
\label{c_derivative}
\end{eqnarray}
where the negative sign corresponds to $\sigma=\uparrow$ 
and the positive sign to $\sigma=\downarrow$. The first 
term on the r.h.s.\ of (\ref{c_derivative}) shows that 
the fermion particles get mixed with the fermion holes. 
The last term in (\ref{c_derivative}) corresponds to 
{\em scattering} of fermions on bosons with finite 
momentum.

As a consequence of (\ref{c_derivative}) we imposed
the following parametrization of the $l$-dependent
fermion operators
\begin{eqnarray}
& & c_{{\bf k}\uparrow}(l)  =  \Pe_{\bf k}(l) \;
c_{{\bf k}\uparrow} + \R_{\bf k}(l) \; 
c_{-{\bf k}\downarrow}^{\dagger}   
\label{c_Ansatz} \\
& + & \; \frac{1}{\sqrt{N}} 
\sum_{{\bf q} \neq{\bf 0}} \left[
p_{{\bf k},{\bf q}}(l) \; b_{\bf q}^{\dagger}
c_{{\bf q}+{\bf k}\uparrow}  + 
r_{{\bf k},{\bf q}}(l) \; b_{\bf q} 
c_{{\bf q}-{\bf k}\downarrow}^{\dagger}
\right] \nonumber \\
& & c_{-{\bf k}\downarrow}^{\dagger}(l) = - \;
\R_{\bf k}^{*}(l) \; c_{{\bf k}\uparrow} + 
\Pe_{\bf k}^{*}(l) \; c_{-{\bf k}\downarrow}^{\dagger} 
\label{cbis_Ansatz} \\ & - &\; \frac{1}{\sqrt{N}} 
\sum_{{\bf q} \neq{\bf 0}} \left[  
r_{{\bf k},{\bf q}}^{*}(l) \; b_{\bf q}^{\dagger}
c_{{\bf q}+{\bf k}\uparrow} -
p_{{\bf k},{\bf q}}^{*}(l) \; b_{\bf q}
c_{{\bf q}-{\bf k}\downarrow}^{\dagger}
\right] . \nonumber 
\end{eqnarray}
In this case, the initial conditions read $\Pe_{\bf k}
(0)=1$, $\R_{\bf k}(0)=0$ and $p_{{\bf k},{\bf q}}(0)=0$,
$r_{{\bf k},{\bf q}}(0)=0$. As we have stated previously \cite{PRL-03}, 
equations (\ref{c_Ansatz},\ref{cbis_Ansatz}) generalize the 
standard Bogoliubov-Valatin transformation \cite{Bogol-Valatin}
in a twofold way:
(a) the initial particle and hole operators are transformed
into cooperons (bogoliubons) through a continuous transformation, 
(b) the scattering of finite momentum Cooper pairs
is additionally taken into account via terms containing
the coefficients $p_{{\bf k},{\bf q}}(l)$ and 
$r_{{\bf k},{\bf q}}(l)$.

>From (\ref{flow}) one can derive the set of differential equations 
for a determination of the $l$-dependent coefficients appearing
in (\ref{c_Ansatz},\ref{cbis_Ansatz}). Because of their importance,
for our further discussion we repeat them here again \cite{PRL-03}
\begin{eqnarray}
 \frac{d\Pe_{\bf k}(l)}{dl} & = &  
\sqrt{n_{cond}^{B}} \; \alpha_{-{\bf k},{\bf k}}(l)
\; \R_{\bf k}(l) , \label{P_flow} \\
& + & \frac{1}{N} \sum_{{\bf q}\neq{\bf 0}} 
\alpha_{{\bf q}-{\bf k},{\bf k}}(l) \left( 
n_{\bf q}^{B} + n_{{\bf q}-{\bf k}\downarrow}^{F}
\right) r_{{\bf k},{\bf q}}(l) 
\nonumber \\ 
\frac{d\R_{\bf k}(l)}{dl} & = & - \;
\sqrt{n_{cond}^{B}} \; \alpha_{{\bf k},-{\bf k}}(l)
\; \Pe_{\bf k}(l) , \label{R_flow} \\
& - & \frac{1}{N} \sum_{{\bf q}\neq{\bf 0}}
\alpha_{-{\bf k},{\bf q}+{\bf k}}(l) \left(
n_{\bf q}^{B} + n_{{\bf q}+{\bf k}\uparrow}^{F}
\right) p_{{\bf k},{\bf q}}(l) 
\nonumber \\ 
\frac{d p_{{\bf k},{\bf q}}(l)}{dl} & = &
\alpha_{-{\bf k},{\bf q}+{\bf k}}(l) \; \R_{\bf k}(l) ,
\label{p_flow} \\
 \frac{d r_{{\bf k},{\bf q}}(l)}{dl} & = & - \;
\alpha_{{\bf k},{\bf q}-{\bf k}}(l)  \Pe_{\bf k}(l) .
\label{r_flow}
\end{eqnarray}
$n_{cond}^{B}$ denotes the concentration of the BE 
condensed bosons and $n_{\bf q}^{B} \equiv \langle 
b_{\bf q}^{\dagger}b_{\bf q} \rangle $ is the distribution 
of finite momentum ${\bf q}\neq{\bf 0}$ bosons.  
Equations (\ref{P_flow}-\ref{r_flow}) properly 
preserve the fermionic anti-commutation relations 
$\left\{ c_{{\bf k}\uparrow}(l),c_{{\bf p}\uparrow}
(l)^{\dagger}\right\} = \delta_{{\bf k},{\bf p}}$ 
and $\left\{ c_{{\bf k}\sigma}(l),c_{{\bf p}\sigma '}
(l)\right\} = 0$.

The single particle fermion Green's function can be easily 
obtained in the limit $l \rightarrow \infty$ when the fermions 
are no longer coupled to the boson subsystem $v_{{\bf k},
{\bf p}}(\infty)=0$. We have shown previously \cite{PRL-03}
that the diagonal part reads 
\begin{eqnarray}
& & A_d^{F}({\bf k},\omega)  \label{f_spectral} \\  
& = & | \Pe_{\bf k}(\infty) |^{2} \delta \left(
\omega - \tilde{\varepsilon}_{\bf k} \right)
+ | \R_{\bf k}(\infty)|^{2} \delta \left( 
\omega + \tilde{\varepsilon}_{-{\bf k}} \right)
\nonumber \\
& + & \frac{1}{N} \sum_{{\bf q}\neq{\bf 0}} 
\left( n_{\bf q}^{B} + n_{{\bf q}+{\bf k}
\uparrow}^{F} \right) | p_{{\bf k},{\bf q}}
(\infty) |^{2} \delta ( \omega + \tilde{E}_{\bf q}
- \tilde{\varepsilon}_{{\bf q}+{\bf k}} ) 
\nonumber \\
& + & \frac{1}{N} \sum_{{\bf q}\neq{\bf 0}}  
\left( n_{\bf q}^{B} + n_{{\bf q}-{\bf k}
\downarrow}^{F} \right) | r_{{\bf k},{\bf q}} 
(\infty) |^{2} \delta ( \omega - \tilde{E}_{\bf q}
+ \tilde{\varepsilon}_{{\bf q}-{\bf k}} ) ,
\nonumber 
\end{eqnarray}
and that it correctly satisfies the sum rule $\int 
d\omega A_d^{F}({\bf k},\omega)=1$. The spectral function 
(\ref{f_spectral}) consists of the delta function peaks which 
represent the long-lived quasi-particles and a remaining 
incoherent background $A_{d,\,inc}^{F}({\bf k},\omega)$ 
of the damped fermion states. It should be mentioned that 
a similar result was obtained by M.\ Fisher and 
coworkers \cite{Lannert-01} for the 2 dimensional Hubbard 
model with a fractionalized structure. Those authors considered 
the spinon and chargon degrees of freedom coupled to a 
Z$_{2}$ gauge field. At low temperatures the spinons and 
chargons are confined. Their deconfinement becomes possible 
at finite temperatures by overcoming the gap of "vison" 
excitations. From an analysis of the low energy excitations 
those authors derived the effective spectral function for physical 
electrons which took exactly the same form as our result 
(\ref{f_spectral}). Apart from a common structure of the spectral 
functions, the remaining discussion and interpretation  
for both models is different.

The spectral function for the off-diagonal part of the fermion 
Green's function $\left< \left< c_{ {\bf k} \uparrow}( \tau );
c_{-{\bf k}\downarrow} \right> \right>$ becomes
\begin{eqnarray}
& & A_{od}^{F}({\bf k},\omega)  \label{off_spectral} 
= \Pe_{\bf k}(\infty)  \R_{\bf k}(\infty) 
\left[\delta \left( \omega + \tilde{\varepsilon}_{\bf k} \right) -
\delta \left( \omega - \tilde{\varepsilon}_{\bf k} \right) \right]
\nonumber \\
& & + \frac{1}{N} \sum_{{\bf q}\neq{\bf 0}} 
p_{{\bf k},{\bf q}}(\infty) r_{{\bf k},{\bf q}}(\infty)
\left\{ \delta ( \omega + \tilde{E}_{\bf q}  - 
\tilde{\varepsilon}_{{\bf q}+{\bf k}} ) 
\right. \label{AF_off} \\  & & \left.  \times
\left[ n_{\bf q}^{B} + n_{{\bf q}+{\bf k}\uparrow}^{F} \right] 
-  \delta ( \omega - \tilde{E}_{\bf q} + 
\tilde{\varepsilon}_{{\bf q}-{\bf k}} ) 
\left[ n_{\bf q}^{B} + n_{{\bf q}-{\bf k}
\downarrow}^{F} \right] \right\} 
\nonumber
\end{eqnarray}
which conserves the sum rule $\int d\omega A^{F}_{od}
({\bf k},\omega)=0$. This function consists also of 
the delta peaks and some fraction distributed over
a wide energy region. In what follows below we shall
study the properties of the diagonal and off-diagonal
spectral functions in various temperature regions.

\subsection{The normal phase spectrum}

Above the critical temperature $T_{c}$ there is 
no boson condensate $n_{cond}^{B}=0$ and in such a case 
the flow equations (\ref{P_flow},\ref{r_flow}) become 
decoupled from (\ref{R_flow},\ref{p_flow}). Consequently  
the coefficients $\R_{\bf k}(l)$ and $p_{{\bf k},{\bf q}}(l)$ 
do not change their initial zero values, i.e.
\begin{eqnarray} 
\R_{\bf k}(l) = 0, \hspace{1cm}
p_{{\bf k},{\bf q}}(l) = 0 .
\label{coeff_aboveTc}
\end{eqnarray}
This property (\ref{coeff_aboveTc}) has a strong influence 
on both fermion spectral functions. The off-diagonal part 
identically vanishes $A^{F}_{od}({\bf k},\omega)=0$ what 
implies that above $T_{c}$ there exists no long range
order parameter in the fermion system.
 
\begin{figure}
\epsfxsize=8cm
\centerline{\epsffile{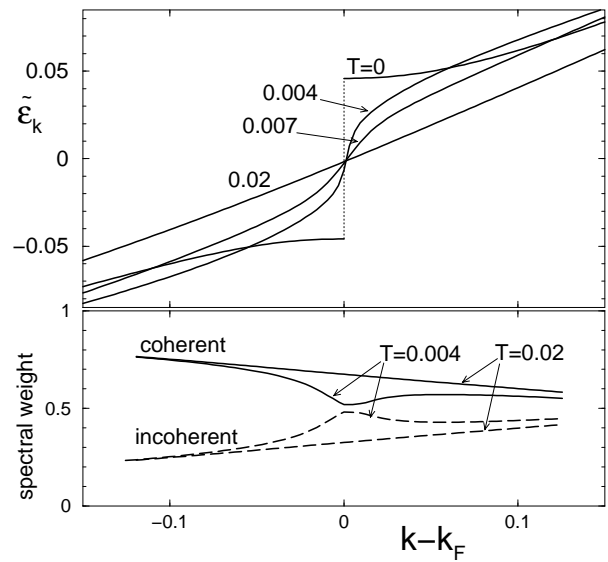}}
\caption{Dispersion of the effective single particle fermion
energy $\tilde{\varepsilon}_{\bf k}$ for several temperatures 
(upper panel) and spectral weights of the coherent and 
incoherent parts for $T=0.02$ and $0.004$ (bottom panel).
Momentum is expressed in units $1/a$.}
\label{Fig4}
\end{figure}

The spectral function of the diagonal part is described by 
\begin{eqnarray} 
& & A_{d}^{F}({\bf k},\omega) = | \Pe_{\bf k}(\infty) |^{2} 
\delta \left( \omega - \tilde{\varepsilon}_{\bf k} \right) 
\label{AF_aboveTc} \\
&+& \frac{1}{N} \sum_{{\bf q}\neq{\bf 0}} \left( n_{\bf q}^{B} 
+ n_{{\bf q}-{\bf k} \downarrow}^{F} \right) 
| r_{{\bf k},{\bf q}} (\infty) |^{2} \delta ( \omega - 
\tilde{E}_{\bf q} + \tilde{\varepsilon}_{{\bf q}-{\bf k}} ).
\nonumber
\end{eqnarray}
We thus have just one branch of long-lived fermion states 
described by the dispersion $\tilde{\varepsilon}_{\bf k}$
and which can be obtained from the renormalization scheme 
discussed by us in Ref.\ \cite{Domanski-01}. For a clear 
understanding of the resulting low energy physics 
we show the temperature evolution of this quantity in figure 
\ref{Fig4} together with the corresponding spectral weight 
$|\Pe_{\bf k}(\infty) |^{2}$. Below a certain characteristic 
temperature $T^{*}$ (in this case $T^{*} \sim 0.1$) we observe 
a tendency to form the pseudogap. Simultaneously 
there occurs a partial transfer of the spectral weight 
from the quasi-particle peak to the incoherent background 
states (see the bottom panel of Fig.\ \ref{Fig4}). 

Such a transfer of the spectral weight is responsible for
the appearance of the Bogoliubov shadow band, as shown from 
the selfconsistent numerical calculations in Ref.\ \cite{PRL-03}.
We will prove this result here analytically. When the temperature
approaches $T_{c}$ from above, bosons tend to occupy macroscopically 
the lowest lying states of energies $\tilde{E}_{\bf q} \simeq 0$. 
These states are spread over the momentum region $|{\bf q}| \leq 
\Lambda$ with the cutoff $\Lambda$ being a fraction of the inverse 
lattice spacing $a^{-1}$. Since the boson occupancy of low momenta 
is  much larger than the fermion population $n^{F}_{{\bf q}-{\bf k}}$ 
we can simplify (\ref{AF_aboveTc}) to the following form
\begin{eqnarray}
A_d^{F}({\bf k},\omega)  & \simeq & | \Pe_{\bf k}(\infty) 
|^{2} \; \delta \left(\omega  - \tilde{\varepsilon}_{\bf k} \right) 
\nonumber \\ & + & 
 | r_{{\bf k},{\bf q} \simeq {\bf 0}}(\infty) |^{2}  
\sum_{|{\bf q}| \leq \Lambda} \frac{n^{B}_{\bf q}}{N} \; 
\delta ( \omega + \tilde{\varepsilon}_{-{\bf k}} - 
\tilde{E}_{\bf q} ) \nonumber \\  
& + &  A^{F}_{d,rigid}({\bf k},\omega) ,
\label{AF_analytic}
\end{eqnarray}
where $A^{F}_{d,rigid}({\bf k},\omega) = N^{-1} \sum_{|{\bf q}| 
> \Lambda} n_{{\bf q}-{\bf k} \downarrow}^{F}  | r_{{\bf k},
{\bf q}} (\infty) |^{2}$ $\times \delta ( \omega + \tilde{
\varepsilon}_{{\bf q}-{\bf k}} - \tilde{E}_{\bf q} )$ 
represents a rigid background which is weakly sensitive to 
varying the temperature. Integrating over the small momentum 
boson states (whose energies are negligible $\tilde{E}
_{\bf q} \simeq 0$) we obtain a partly broadened peak 
with its maximum occurring at $\omega=-\tilde{\varepsilon}
_{\bf k}$. This additional branch of fermion excitations 
can well be fitted by the Lorentzian shape
\begin{eqnarray}
\sum_{|{\bf q}| \leq \Lambda} \frac{n^{B}_{\bf q}}{N} \; 
\delta ( \omega + \tilde{\varepsilon}_{-{\bf k}} - 
\tilde{E}_{\bf q} ) \simeq \frac{n^{B}}{\pi} \;
\frac{\Gamma_{\bf k}}{ ( \omega + \tilde{\varepsilon}
_{-{\bf k}} )^{2} + \Gamma_{\bf k}^{2}}
\label{lorenzian}
\end{eqnarray}
with $n^{B}$ being the  concentration of bosons. For $T<T_{c}$ 
(\ref{lorenzian}) shrinks to $\Gamma \rightarrow 0$. Similar 
Bogoliubov shadow bands were also indicated for the normal 
phase of the $U<0$ Hubbard model using the Quantum Monte Carlo 
studies \cite{Zurich_group} and the conserving diagrammatic 
treatment of Tremblay and coworkers \cite{Tremblay-97}. In 
our case the Bogoliubov-like branches are characterized by
the pseudogap dispersion $\pm \tilde{\varepsilon}_{\bf k}$
shown here in figure 4. Above $T_{c}$ the shadow branch 
(corresponding to $-$ sign) is broad as discussed in our 
Letter \cite{PRL-03}. 

\begin{figure}
\epsfxsize=8.5cm
\centerline{\epsffile{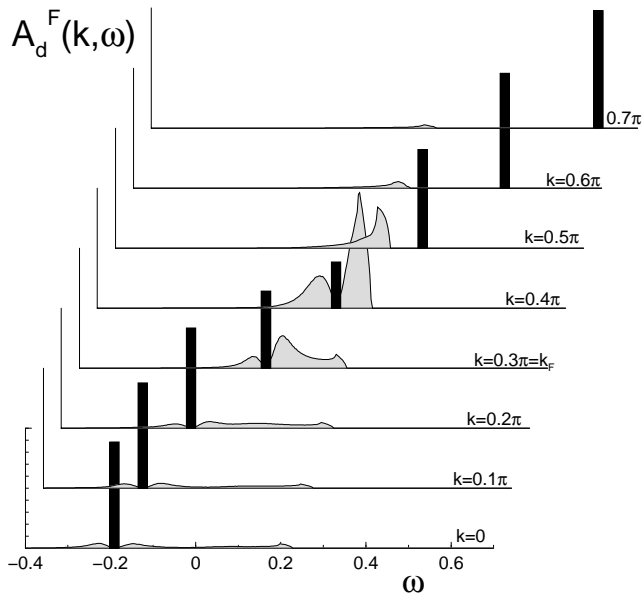}}
\caption{Diagonal part of the single particle fermion 
spectral function (\ref{AF_aboveTc}) for a large part of 
the Brillouin zone at $T=0.007$. Vertical bars represent 
the coherent part $A_{d,coh}^{F}({\bf k},\omega)$ while 
the shaded area corresponds to $A_{d,inc}^{F}({\bf k},
\omega)$. Besides only a close vicinity of the Fermi 
surface (consult (\ref{lorenzian}) or see the figure 1 in 
\cite{PRL-03}) this spectral function is rather weakly 
affected by temperature.}
\label{Fig5}
\end{figure}

Far away from the Fermi surface the renormalized energies are 
nearly equal to the bare values $\tilde{\varepsilon}_{\bf k}
=\varepsilon_{\bf k}-\mu$. The long-lived quasi-particle 
contains however only part of the spectral weight, i.e.,  
$|\Pe_{\bf k}(\infty)|^{2}<1$. In figure \ref{Fig5} 
we illustrate the coherent and incoherent contributions 
of the single particle fermion spectral function 
(\ref{AF_aboveTc}) in a large part of the Brillouin zone.
The damped fermion states are spread over the large energy 
regime $\omega$, they exist even for momenta quite distant 
from $k_{F}$. This result agrees well with the previous 
studies based on  selfconsistent perturbative theory 
(see the third reference of \cite{Ranninger_etal}). The presence 
of the substantial incoherent background states might possibly 
be related to the experimental signal observed in the ARPES 
measurements for the HTS compounds \cite{ARPES-exp}.

\vspace{0.5cm}
\subsection{The long-lived quasiparticles below $T_{c}$}

Below the critical temperature $T_{c}$ there exists a finite 
amount of the BE condensed bosons $n_{cond}^{B} \neq 0$. 
This has important effects on the fermionic spectrum. 
Using the flow equation method we have shown analytically 
\cite{Domanski-01} that $\tilde{\varepsilon}_{{\bf k}} 
= \mbox{sgn}(\varepsilon_{\bf k}-\mu) \sqrt{
(\varepsilon_{\bf k}-\mu)^{2}+v^{2}n_{cond}^{B}}$. 
In the superconducting phase the effective dispersion is 
thus characterized by the true gap $\Delta_{sc}(T)=
v \sqrt{n_{cond}^{B}(T)}$.

For a finite condensate fraction $n_{cond}^{B}$, we can see 
from equations (\ref{R_flow},\ref{p_flow}) that the coefficients 
$\R_{\bf k}$ and $p_{{\bf k},{\bf q}}$ become finite too. 
The long-lived fermion excitations are then represented 
by the two sharp Bogoliubov branches
\begin{eqnarray}
A_{d,coh}^{F}({\bf k},\omega) & = &
 | \Pe_{\bf k}(\infty) |^{2} \; \delta \left(
\omega - \tilde{\varepsilon}_{\bf k} \right)
\nonumber \\
& + & | \R_{\bf k}(\infty)|^{2} \; \delta \left(
\omega + \tilde{\varepsilon}_{{\bf k}} \right) .
\label{diag_coh}
\end{eqnarray}
Similarly, the coherent part of the off-diagonal spectral 
function (\ref{off_spectral}) reads
\begin{eqnarray}
A^{F}_{od,coh}({\bf k},\omega) & = &
\Pe_{\bf k}(\infty)  \R_{\bf k}(\infty) \nonumber \\
& \times & \left[\delta \left( \omega + \tilde{
\varepsilon}_{\bf k} \right) - \delta \left( \omega 
- \tilde{\varepsilon}_{\bf k} \right) \right] .
\label{offdiag_coh}
\end{eqnarray}
Equations (\ref{diag_coh},\ref{offdiag_coh}) resemble 
the BCS-type results which usually appear in the mean 
field studies of this model \cite{Micnas_etal,Lee_etal,
Ranninger_etal}. Let us stress that in distinction to 
these we obtain here the total spectral weight engaged 
in the coherent parts to be \cite{PRL-03}
\begin{eqnarray}
| \Pe_{\bf k}(\infty) |^{2}+|\R_{\bf k}(\infty) 
|^{2} < 1 . 
\end{eqnarray}
The rest of the weight is redistributed over  the damped 
fermionic states. This will be discussed separately 
in the next section. 

In order to show the correspondence of our present 
analysis to the previous results for this BF model 
we prove in Appendix A, that upon neglecting the incoherent 
background states of (\ref{f_spectral},\ref{off_spectral})
one exactly obtains the usual BCS coherence factors
\begin{eqnarray}  
| \Pe_{\bf k}(\infty) |^{2} = \frac{1}{2} \left( 1 +
\frac{| \varepsilon_{\bf k}-\mu |}{\xi_{\bf k}^{MF}}
\right) & = & 1 - | \R_{\bf k}(\infty) |^{2}  ,
\label{BCS_factors} \\
\Pe_{\bf k}(\infty) \; \R_{\bf k}(\infty) 
& = & - \; \frac{v\sqrt{n_{cond}^{B}}}
{2\xi_{\bf k}^{MF}} ,
\label{BCS_off}
\end{eqnarray}
where $\xi_{\bf k}^{MF}=\sqrt{(\varepsilon_{\bf k}
-\mu)^{2}+\Delta_{sc}^{2}}$. By  neglecting  the 
incoherent parts, the flow equation procedure 
reproduces exactly the standard mean field equations
\cite{Micnas_etal,Lee_etal,Ranninger_etal}
\begin{eqnarray} 
\langle c_{{\bf k}\uparrow}^{\dagger}
c_{{\bf k}\uparrow} \rangle & = &
\frac{1}{2} \left[ 1 - \frac{\varepsilon_{\bf k}-\mu}
{ \xi_{\bf k}^{MF}} \; \mbox{tgh}\left(
\frac{\xi_{\bf k}^{MF}}{2k_{B}T} \right) \right] ,
\label{conc_MF} \\
\langle c_{-{\bf k}\downarrow}c_{{\bf k}\uparrow} 
\rangle & = & - \; \frac{v \; n_{cond}^{B}}
{2\xi_{\bf k}^{MF}} \; \mbox{tgh}\left(
\frac{\xi_{\bf k}^{MF}}{2k_{B}T} \right) .
\label{gap_MF}
\end{eqnarray}
We shall generalize these equations (\ref{conc_MF},
\ref{gap_MF}) in the next section, taking into account the 
contribution from the damped fermion states.

\subsection{Damped quasi-particles below $T_{c}$}  

The finite life-time (damped) states do participate in the 
fermionic excitation spectrum both above $T_{c}$ as well as
below of it. Nevertheless their presence does not spoil 
a long range coherent behavior between the fermion pairs 
which is necessary for superconductivity to occur. In chapter 
III we saw that boson as well as the fermion pair spectra 
are characterized below $T_{c}$ by  long-lived collective
modes with  $\tilde{E}_{\bf q} \propto |{\bf q}|$ 
\cite{Domanski-01} which becomes separated from the 
surrounding background of incoherent states by 
the gap $2\Delta_{sc}$. We show below that in the single 
particle excitation spectrum the incoherent states can 
exist only outside the gap $\Delta_{sc}$ around 
the Fermi energy.  

The contribution from the damped fermionic states is 
described by the following part of the spectral 
function (\ref{f_spectral}) 
\begin{eqnarray}
& & A_{d,inc}^{F}({\bf k},\omega)  =  
A_{d}^{F}({\bf k},\omega) - A_{d,coh}^{F}({\bf k},\omega)
\nonumber \\
& = &\frac{1}{N} \sum_{{\bf q}\neq{\bf 0}} \left[
\left( n_{\bf q}^{B} + n_{{\bf q}+{\bf k}
\uparrow}^{F} \right) | p_{{\bf k},{\bf q}}
(\infty) |^{2} 
\delta ( \omega + \tilde{E}_{\bf q}
- \tilde{\varepsilon}_{{\bf q}+{\bf k}} )
\right. \nonumber \\  & + & \left.  
\left( n_{\bf q}^{B} + n_{{\bf q}-{\bf k}
\downarrow}^{F} \right) | r_{{\bf k},{\bf q}}
(\infty) |^{2} \delta ( \omega - \tilde{E}_{\bf q}
+ \tilde{\varepsilon}_{{\bf q}-{\bf k}} ) \right] 
\label{A_incoh}
\end{eqnarray}
and similarly in case of the off-diagonal function 
(\ref{off_spectral})
\begin{eqnarray}
& & A_{od,inc}^{F}({\bf k},\omega)  =  
- \; \frac{1}{N} \sum_{{\bf q}\neq{\bf 0}} 
p_{{\bf k},{\bf q}}(\infty) 
r_{{\bf k},{\bf q}}(\infty) 
\nonumber \\ & & \times 
\left[ \left( n_{\bf q}^{B} + n_{{\bf q}+{\bf k}
\uparrow}^{F} \right) \delta ( \omega + \tilde{E}_{\bf q}
- \tilde{\varepsilon}_{{\bf q}+{\bf k}} )
\right. \nonumber \\  & & \left.  - 
\left( n_{\bf q}^{B} + n_{{\bf q}-{\bf k}
\downarrow}^{F} \right) \delta ( \omega - \tilde{E}_{\bf q}
+ \tilde{\varepsilon}_{{\bf q}-{\bf k}} ) \right] .
\label{A_off_incoh}
\end{eqnarray}

It is instructive to consider first the incoherent 
spectral functions (\ref{A_incoh},\ref{A_off_incoh}) for 
the ground state $T=0$. Finite momenta bosons have 
$\tilde{E}_{{\bf q}\neq{\bf 0}}>0$, such that all bosons 
are condensed at $T=0$ and we can put $n_{\bf q}^{B}=0$ 
for any ${\bf q}\neq{\bf 0}$. On the other hand, 
fermions occupy only the states below the Fermi 
surface, i.e.\ $n_{{\bf k}\sigma}^{F}=\theta(-\tilde
{\varepsilon}_{\bf k})n_{{\bf k}\sigma}^{F}$. Therefore, 
the first term on the r.h.s.\ of (\ref{A_incoh}) gives 
rise to the appearance of the incoherent fermion states 
at negative energies. It can be written as
\begin{eqnarray}  
& & A_{d,inc}^{F}({\bf k},\omega < 0)  \label{inc_neg} \\ & = & 
\frac{1}{N} \sum_{\tilde{\varepsilon}_{{\bf q}+{\bf k}}<0} 
n^{F}_{{\bf q}+{\bf k} \uparrow } | p_{{\bf k},{\bf q}} 
(\infty) |^{2} \delta \left[ \omega + \left( \tilde{E}_{\bf q} +  
| \tilde{\varepsilon}_{{\bf q}+{\bf k}} | \right) \right]. 
\nonumber
\end{eqnarray}
Similarly, the last term in (\ref{A_incoh}), which
corresponds to incoherent states at positive energies,
becomes
\begin{eqnarray}  
& & A_{d,inc}^{F}({\bf k},\omega > 0)  \label{inc_pos} \\ & = &  
\frac{1}{N} \sum_{\tilde{\varepsilon}_{{\bf q}-{\bf k}}<0} 
n^{F}_{{\bf q}-{\bf k} \downarrow } | r_{{\bf k},{\bf q}}
(\infty) |^{2} \delta \left[ \omega - \left( \tilde{E}_{\bf q} + 
| \tilde{\varepsilon}_{{\bf q}-{\bf k}}| \right) \right] .
\nonumber
\end{eqnarray}
The off-diagonal spectral function (\ref{A_off_incoh})
can be derived in the same manner. From inspecting 
their structure we notice that: 
\begin{itemize} 
\item{no damped fermion states are allowed to occur 
    within the superconducting energy gap window 
    $|\omega| \leq \Delta_{sc}$ because 
    $|\tilde{\varepsilon}_{{\bf q}\pm{\bf k}}| 
    \geq \Delta_{sc}$,} 
\item{even outside the superconducting gap, in a close 
    vicinity of the coherence peaks (\ref{diag_coh})  
    there is a partial suppression of the damped 
    fermion states due to finite $\tilde{E}_{\bf q}$
    in equations (\ref{inc_neg},\ref{inc_pos}). 
    As shown in our earlier studies 
    \cite{Ranninger_etal,Domanski-01} the width of 
    such a boson band is rather small and of the order 
    $v^{2}/D$.}
\end{itemize}
The incoherent fermion states can thus be formed only 
outside the superconducting gap $|\omega|>\Delta_{sc}$
and yet they are strongly suppressed over some small
energy region (of the order $v^{2}/D$ from the coherence 
peaks).

To summarize our results of the last two sections ,
we conclude that the single particle fermion spectral 
function (\ref{f_spectral}) is characterized in the 
superconducting phase by: 
(a) the presence of  narrow quasi-particle {\em peaks}
    at $\omega=\pm \sqrt{(\varepsilon_{\bf k}-\mu)^{2}
    +\Delta_{sc}^{2}}$,
(b) the partial suppression of the fermionic states 
    very close to the coherence peaks, giving rise 
    to the appearance of the {\em dip}-like structure,  
(c) the existence of a broad incoherent background 
    spectrum with its flat maximum ({\em hump}) 
    situated fairly away from the Fermi energy.
Upon increasing the temperature we expect the {\em dip}-like  
structure to be gradually filled in via absorbing part of the spectral 
weight from the coherent Bogoliubov 
peaks. Our expectation is motivated here by the fact 
that the spectral weights $|\Pe_{\bf k}(\infty)|^{2}$, 
$|\R_{\bf k}(\infty)|^{2}$ of the coherent peaks 
are very sensitive to temperature because of 
$n^{B}_{cond}(T)$ appearing in their flow equations 
(\ref{P_flow},\ref{R_flow}).

\begin{figure}
\epsfxsize=8cm
\centerline{\epsffile{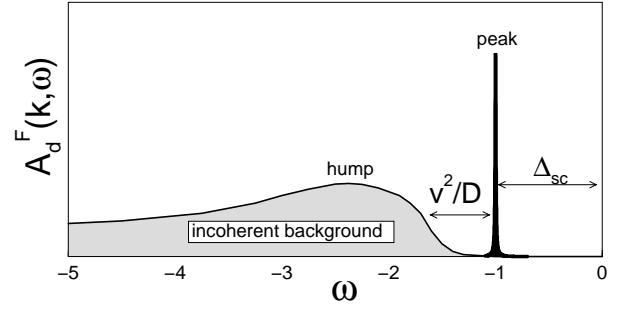}}
\caption{Schematic view of the single particle 
fermion spectral function $A_d^{F}({\bf k}_{F},\omega<0)$ 
in the ground state of the boson fermion model. Energy
$\omega$ is expressed in units of the superconducting
gap which is roughly scaled via $\Delta_{sc} \simeq 
v \sqrt{n^{B}_{cond}}$.}
\label{Fig6}
\end{figure}

Our theoretical prediction concerning the single 
particle spectrum of the superconducting phase is in 
qualitative agreement with the experimental data 
known for the high $T_{c}$ underdoped cuprates. The 
direct photoemission spectroscopy probes the occupied 
spectrum $A^{F}({\bf k},\omega)\times \left[ 1+
\mbox{exp}\left(\omega/k_{B}T\right) \right]^{-1}$. 
Photoemission measurements indeed revealed the appearance
of such {\em peak} - {\em dip} - {\em hump} structure
\cite{ARPES-exp}. This issue has been so far interpreted 
in terms of the boson resonant mode to which the single 
particle excitations are coupled. Several candidates
have been proposed in the literature for such boson 
modes \cite{Schrieffer,Chubukov,Eschrig}. The BF model 
(\ref{BF}) is an alternative scenario, where bosons 
represent nearly localized electron pairs.

Concluding this discussion on the coherent and 
incoherent single particle spectra, let us stress that the 
expectation values for the particle distribution
$n^{F}_{\bf k \sigma}$ at arbitrary temperature can be 
obtained by integrating over the expression (\ref{f_spectral}) 
after having been multiplied by the Fermi Dirac function 
$f(x)=\left( 1+e^{x/k_{B}T} \right)^{-1}$, i.e.\ 
\begin{eqnarray}
n^{F}_{{\bf k} \sigma} = \int_{-\infty}^{\infty} d\omega \;
A_d^{F}({\bf k},\omega) \; f(\omega). \label{distrib_general} 
\end{eqnarray}
Similarly,  the order parameter determined via (\ref{off_spectral}) 
is given by 
\begin{eqnarray} 
& & \langle c_{-{\bf k}\downarrow}c_{{\bf k}\uparrow} \rangle 
= \Pe_{\bf k}(\infty) \R_{\bf k}(\infty) \left[ 1 - 2 
f(\tilde{\varepsilon}_{\bf k}) \right] \nonumber \\
&& + \; \frac{1}{N} \sum_{{\bf q}\neq{\bf 0}} 
p_{{\bf k},{\bf q}}(\infty) r_{{\bf k},{\bf q}}(\infty) 
\left[ \left( n_{\bf q}^{B} + n_{{\bf q}+{\bf k}
\uparrow}^{F} \right) \right. \label{gap_general} 
\\ && \times \left. f(\tilde{\varepsilon}_{{\bf q}+{\bf k}} 
-  \tilde{E}_{\bf q} )-\left( n_{\bf q}^{B} + n_{{\bf q}-{\bf k}
\downarrow}^{F} \right) f( - \tilde{\varepsilon}_{{\bf q}
-{\bf k}} +  \tilde{E}_{\bf q} ) \right] .
\nonumber 
\end{eqnarray}
which generalizes the mean field equation (\ref{gap_MF}).
Unfortunately, at the present stage we are not able 
to solve numerically these equations (\ref{distrib_general},
\ref{gap_general}) for some realistic $d>2$ system when 
$T_{c}>0$. We will try to address this problem in the 
future by applying some approximate treatment.

\section{Conclusions}

In the present paper we studied the mutual relations 
between the single particle and the pair excitation spectra 
within the BF model (\ref{BF}). We investigated their 
interdependence occurring in the superconducting phase 
below the critical temperature, and in the pseudogap 
region above $T_{c}$. Many-Body effects were treated 
by means of the continuous canonical transformation 
\cite{Wegner-94} originating from a general framework 
of the renormalization group technique \cite{RG-idea}. 

We have earlier reported \cite{Domanski-01} that, upon 
approaching $T_{c}$ from above, there is a partial suppression 
of the single particle fermion states near the Fermi energy 
(pseudogap). Here, we supplement this picture by showing 
that the pseudogap feature is accompanied by a subsequent 
emergence of the fermion pair properties. Such pairs can 
show up in the pseudogap region as long lived entities 
provided that their total momentum is finite and larger 
than a certain $q_{crit}(T)$. For $T \rightarrow T_{c}$ the 
critical momentum is $q_{crit}=0$, therefore all the fermion 
pairs become good quasi-particles. The zero momentum fermion 
pairs exist above $T_{c}$ only as the damped objects
because of their overlap with the incoherent background. 
However, upon lowering the temperature such an overlap 
gradually diminishes which effectively leads to increase  
their life-time. 

Although above $T_{c}$ there exist no off-diagonal 
long range order (ODLRO) we predict  nevertheless the 
possibility to observe some ordering on a 
restricted spatial and temporal scale. Experiments 
using terahertz frequencies 
\cite{Corson-99} did indeed confirm the existence of 
preformed pairs in the underdoped HTS materials 
up to 25 degrees above $T_{c}$. We moreover suspect, 
that  "moving" fermion pairs (on the basis of our
study expected to have the infinite life-time) have 
been observed in the measurements of the Nernst 
coefficient \cite{Ong-00}. 

As far as the single particle excitations are concerned
we provided here the analytical arguments for the appearance 
of Bogoliubov-type bands in the pseudogap phase. Upon 
approaching $T_{c}$ from above the shadow branch absorbs 
more an more spectral weight and simultaneously narrows 
as was previously indicated by us from the selfconsistent 
numerical study \cite{PRL-03}. Below $T_{c}$, the shadow 
branch shrinks to the usual delta function peak and  marks the 
appearance of infinite life-time cooperons. In distinction 
to the conventional BCS superconductors we  find 
that the quasi-particle peaks occurring 
at $\omega=\pm \sqrt{(\varepsilon_{\bf k}-\mu)^{2}+
\Delta_{sc}^{2}}$ become slightly separated from the rest 
of the spectrum which is present in a form of the incoherent 
background. This effectively leads to the formation of the 
intriguing {\em peak} - {\em dip} - {\em hump} structure 
which has been well documented experimentally by  ARPES 
measurements. In our scenario such a structure is a consequence 
of the pair correlations (see also the similar conclusion 
in the Ref.\ \cite{Pieri-04}).

In a future study we plan take into 
consideration an anisotropic 2-dimensional version of 
this model which is more realistic for describing the 
HTS cuprates. The boson-fermion coupling $v$ should then  
be appropriately factorized by the $d$-wave form factor. 
Another important aspect which has not been fully explored so far 
concerns the doping dependence of the pseudogap and other related 
precursor features discussed above. Such a study requires 
a proper treatment of the hard-core nature of the local 
electron pairs, which is an issue which is rather nontrivial. 
Preliminary results were so far obtained using 
a perturbative approach \cite{Tripodi-03} and from  
exact diagonalization studies \cite{Cuoco-03}.

\section{Acknowledgment}
We kindly acknowledge instructive discussions with 
Prof.\ F.\ Wegner on the flow equation procedure.
T.D.\ moreover acknowledges a partial support from 
the Polish Committee of Scientific Research under 
the grant No.\ 2P03B06225.

\appendix*
\section{}

Neglecting the incoherent background states of the 
single particle functions $A_{d}^{F}({\bf k},
\omega)$ and $A_{od}^{F}({\bf k},\omega)$ is 
equivalent to the assumption that $p_{{\bf k},
{\bf q}}(l)=0$ and $r_{{\bf k},{\bf q}}(l)=0$.
In such case the flow equations (\ref{P_flow},
\ref{R_flow}) simplify to
\begin{eqnarray} 
 \frac{d\Pe_{\bf k}(l)}{dl} & = &
\sqrt{n_{cond}^{B}} \; \alpha_{-{\bf k},{\bf k}}(l)
\; \R_{\bf k}(l) \label{A1} \\
\frac{d\R_{\bf k}(l)}{dl} & = & - \;
\sqrt{n_{cond}^{B}} \; \alpha_{{\bf k},-{\bf k}}(l)
\; \Pe_{\bf k}(l) \label{A2} .
\end{eqnarray}
We will now assume that both functions 
$\Pe_{\bf k}(l)$ and $\R_{\bf k}(l)$ are real
(this requirement does not restrict the generality
of our considerations). We rewrite (\ref{A1}) as
\begin{eqnarray}
 \frac{d\Pe_{\bf k}(l)}{R_{\bf k}(l)} =
\sqrt{n_{cond}^{B}} \; \alpha_{-{\bf k},
{\bf k}}(l) dl 
\label{A3}
\end{eqnarray}
and integrate  both sides of (\ref{A3})
in the limits $\int_{l=0}^{l=\infty}$. Using the
sum rule (10) of Ref.\ \cite{PRL-03} we can substitute
$\R_{\bf k}(l)=\sqrt{1-\Pe_{\bf k}^{2}(l)}$.
Upon integration we get for the l.h.s.\
\begin{eqnarray} 
\int_{l=0}^{l=\infty} \frac{d\Pe_{\bf k}(l)}
{\sqrt{1-(\Pe_{\bf k}(l))^{2}}} = - \;
\mbox{arcos} \left[ \Pe_{\bf k}(\infty) \right]
\label{A4}
\end{eqnarray}
because $\mbox{arcos}\left[ \Pe_{\bf k}(0) 
\right] =0$.
                          
Integration of the r.h.s.\ of (\ref{A3}) requires
the knowledge of $\alpha_{-{\bf k},{\bf k}}(l)$
for the superconducting phase. First, let us notice that 
from the general definition of this parameter we
have $\alpha_{-{\bf k},{\bf k}}(l) = 2 \left[ 
\varepsilon_{\bf k}(l) - \mu \right] 
v_{-{\bf k},{\bf k}}(l).$ Using the equation 
(43) of the Ref.\ \cite{Domanski-01} we can
further write 
\begin{eqnarray}  
v_{-{\bf k},{\bf k}}(l) \; dl = -\; \frac{ d \;
v_{-{\bf k},{\bf k}}(l)}{4 \left[ \varepsilon_{\bf k}
(l)-\mu \right] }  .
\label{A5}
\end{eqnarray}
In the superconducting phase we can additionally 
make use of the invariance shown in the equation 
(51) of the Ref.\ \cite{Domanski-01}
\begin{eqnarray}
\varepsilon_{\bf k}(l)-\mu = \pm
 \sqrt{ ( \xi_{\bf k}^{MF})^{2} - n_{cond}^{B} 
v_{-{\bf k},{\bf k}}^{2}(l)}
\label{A6}
\end{eqnarray}
where $\pm = \mbox{sgn}(\varepsilon_{\bf k}-\mu)$.
By substituting (\ref{A5}) and (\ref{A6}) into 
the r.h.s.\ of equation (\ref{A4}) we obtain  
\begin{eqnarray}
\sqrt{n_{cond}^{B}}  \int_{0}^{\infty} & dl &
\alpha_{-{\bf k},{\bf k}}(l)   =  - 
\frac{\sqrt{n_{cond}^{B}}}{2} \int_{l=0}^{l=\infty}
\frac{ d \; v_{-{\bf k},{\bf k}}(l)} {
\varepsilon_{\bf k}(l)-\mu  }
\nonumber \\ & & \nonumber \\
& = & \frac{\mp 1}{2} 
\sqrt{n_{cond}^{B}} \int_{v}^{0}
\frac{ d \; v_{-{\bf k},{\bf k}}} {\sqrt{
(\xi_{\bf k}^{MF})^{2} - n_{cond}^{B} 
v_{-{\bf k},{\bf k}}^{2}}}
\nonumber \\ 
& = &\pm  \frac{1}{2}  
\left[ \frac{\pi}{2} - \mbox{arcos}\left(
\frac{v\sqrt{n_{cond}^{B}}}{\xi_{\bf k}^{MF}}
\right) \right] .
\label{A7}
\end{eqnarray}
In the second line of (\ref{A7}) we applied the initial
condition $v_{-{\bf k},{\bf k}}(0)=v$ and also the final
result $v_{-{\bf k},{\bf k}}(\infty)=0$ when changing the 
integration variable from $dl$ to $dv_{-{\bf k},{\bf k}}(l)$.

By comparing the results (\ref{A4}) and (\ref{A7})
we obtain 
\begin{eqnarray}  
\mp 2 \; \mbox{arcos} \left[ \Pe_{\bf k}(\infty) 
\right] =  \frac{\pi}{2} - \mbox{arcos}\left( 
\frac{v\sqrt{n_{cond}^{B}}}{\xi_{\bf k}^{MF}}
\right) 
\label{A8}
\end{eqnarray}
and by taking the cosine function on both sides
we get $2 \Pe_{\bf k}^{2}(\infty) - 1 = 
| \varepsilon_{\bf k}-\mu | / \xi_{\bf k}^{MF}$
which leads to the usual BCS coherence factors
\begin{eqnarray}  
\Pe_{\bf k}^{2}(\infty)  = \frac{1}{2} \left[ 
1 + \frac{ | \varepsilon_{\bf k}-\mu | }
{\xi_{\bf k}^{MF}} \right] = 1 - 
\R_{\bf k}^{2}(\infty) .
\label{A9}
\end{eqnarray}
In this way we also know that the magnitude
of the product is $| \Pe_{\bf k}(\infty) 
\R_{\bf k}(\infty)| = v\sqrt{n_{cond}^{B}}/
2\xi_{\bf k}^{MF}$. Hence, assuming that 
$\Pe_{\bf k}(l)$ is positive (in particular 
also for $l=\infty$) then on a basis of the  
flow equation (\ref{A2}) we conclude that 
$\R_{\bf k}(l) =-\mbox{sgn}(\varepsilon_{\bf k}
-\mu) |\R_{\bf k}(l)|$ and thus finally 
\begin{eqnarray} 
\Pe_{\bf k}(\infty) \; \R_{\bf k}(\infty) 
= - \; \frac{v\sqrt{n_{cond}^{B}}}
{2\xi_{\bf k}^{MF}} .
\label{A10}
\end{eqnarray}
This product (\ref{A10}) enters the off-diagonal
Green's function and in consequence yields the equation
(\ref{gap_MF}) for the order parameter.

\end{document}